\documentclass[aps,pre,twocolumn]{revtex4}
\usepackage{amsmath,bm}
\usepackage{feyn}

\begin{document}

\title{Non-linear fluctuation effects in dynamics of freely suspended film}

\author{E.I.Kats$^{1,2}$, V.V.Lebedev$^{1,2}$ 
}

\affiliation{$^1$Landau Institute for Theoretical Physics, RAS, \\
142432, Chernogolovka, Moscow region, Russia \\
$^2$Moscow Institute of Physics and Technology, \\
141700, Dolgoprudny, Moscow region, Russia.
}

\begin{abstract}

Long-scale dynamic fluctuation phenomena in freely suspended films is analyzed. We consider isotropic films that, say, can be pulled from bulk smectic $A$ liquid crystals. The key feature of such objects is possibility of bending deformations of the film. The bending (also known as flexular) mode turns out to be anomalously weakly attenuated. In the harmonic approximation there is no viscous-like damping of the bending mode, proportional to $q^2$ ($q$ is the wave vector of the mode), since it is forbidden by the rotational symmetry. Therefore the bending mode is strongly affected  by non-linear dynamic fluctuation effects. We calculate the dominant fluctuation contributions to the damping of the bending mode due to its coupling to the in-plane viscous mode, that restores the viscous-like $q^2$ damping of the bending mode. Our calculations are performed in the framework of the perturbation theory where the coupling of the modes is assumed to be small, then the bending mode damping is relatively weak. We discuss our results in the context of existing experiments and numeric simulations of the freely suspended films and propose possible experimental observations of our predictions.

\end{abstract}


\maketitle

\section{Introduction}
\label{sec:intro}

Thin flexible films and membranes (e.g., lipid bilayers) are curious but ubiquitous objects in the realm of condensed matter science. In the long-scale limit they can be considered as two-dimensional objects embedded into three-dimensional space. The main peculiarity of such objects is possibility of their bending distortions that have to be analyzed in addition to traditional long-scale deformations like density variations and in-plane motions. Here we turn our attention to freely suspended films that are not surrounded by a dense matter (unlike the biological membranes), that makes their bending oscillations weakly attenuating and, as a result, sensitive to non-linear fluctuation effects. While basic thermal (and even quantum) fluctuation effects in various three- and two- dimensional condensed matter systems are well studied and known, the situation with freely suspended films is not completely recognized. In the paper we analyze liquid-like films that are isotropic thin sheets of matter.

The very existence of freely suspended films, technology of their production and basic experimental and theoretical studies are known from 70-ties of the previous century, and summarized in a number of review publications and monographs (see, to quote some of them,
\cite{NP89,CL00,BT01,SA03,JO03,DK14}). Majority of experimental and theoretical investigations of physical properties of the free standing liquid-crystalline films and fluid membranes are devoted to thermodynamic and structural characteristics. In terms of correlation functions of long-scale degrees of freedom, it corresponds to measurements, numeric simulations or calculations of the static structure factor. Some works were performed also to study dynamic structure factor or time dependent correlation functions (see cited above literature and references therein), and only a very few results concerning non-linear dynamics and fluctuation effects are known \cite{KL85,LM89,KL93}. However, in all those works the film was assumed either to have a finite (and not too small) surface tension, or (for tensionless membrane) to be surrounded by viscous isotropic liquid.

In a parallel world of crystalline or glassy freely supported membranes investigations of the vibrational dynamics also started long ago from the very influential works \cite{NP87,AL88}. In these and some other papers (see also the review \cite{BT01}) in contrast to fluid-like films another situation has been considered. Namely, solid tensionless membranes freely suspended in the vacuum (or in a dilute gas) were examined. These types of the investigations were resurrected recently (see e.g., \cite{NJ05,MO08,ZR10,LK12,KL14}) with a motivation (besides own fundamentally  interesting physics) of various electro-opto-mechanical applications of the graphene films. However, fluctuation effects in solid and in liquid films are dramatically different. One can easily see the difference even in static properties. Indeed, as it is well known (compare results for solid films \cite{NP87} and fluid membranes \cite{HE73} or in \cite{SA03}), thermal fluctuations make the freely suspended solid membrane more rigid, whereas for liquid membranes thermal fluctuations make the membrane more floppy. This qualitatively different behavior is accounted for the different in the both cases (solid and fluid membranes) sources of the anharmonic terms, responsible for the bending fluctuations. In the former case the principal anharmonic terms come from the coupling of in-plane and out-of-plane (flexular) degrees of freedom, whereas in the latter case the bending softening occurs as a result of the intrinsic pure bending anharmonicity. Summarizing this brief overview of the state of art with film vibrational fluctuations, we conclude that there are some studies of fluid membranes surrounded by a viscous isotropic liquid and/or under tension, but the case of tensionless freely suspended in vacuum (or gas) membranes has been investigated only for solid membranes. The purpose of our work is to close this problem overlooked in the previous works.

An interest to fluctuation dynamic phenomena reflects both their practical importance and related fundamental science challenges. In this paper we extend the previous works, see Refs. \cite{NP89,CL00,BT01,SA03,JO03,DK14}, and also \cite{KL93}, to study dynamic fluctuation effects in the freely suspended in vacuum (or gas) tensionless liquid films. As experiments grow in resolution and in sophistication, it is likely that further details of dynamics will be revealed whose interpretation will require a reliable theoretical approach, we describe in our work. It is worth to note to the point that structurally similar to fluid films, polymerized or amorphous glassy membranes dynamically belong to the class of crystalline membranes (see, e.g., \cite{LK12}). Indeed, in such membranes the in-plane overdamped viscous mode (whose coupling with the bending mode is the main source of fluctuation contributions into the mode damping) is replaced by the propagating transverse sound mode.

Our paper is structured as follows. In the next section \ref{sec:basic}, we introduce main ingredients of our model, define physical quantities of interest, and derive basic relations for the quantities. In section \ref{sec:action} we introduce the key tool for theoretical study of dynamic fluctuations -- the effective action. Then in section \ref{sec:perturbation} we give details for the perturbative treatment of the effective action functional, and formulate our main results in this work, namely the pure fluctuation bending mode damping and fluctuation contribution into the in-plane shear viscosity. In Conclusion we present a summary of our results. Appendix contains some details of the calculations.

 \section{Basic Relations}
 \label{sec:basic}

We examine physical properties of freely suspended thin films. Very thin films (consisting of a few molecular layers) can be pulled from bulk smectic liquid crystal phases \cite{CH92,GP93,JO03}. It should be borne in mind that stability of such thin films in the freely suspended state is accounted for by the layered structure inherent in all smectics, i.e., by the presence of some internal interaction forcing molecules to stay in one layer. At increasing temperature, when this layered structure is destroyed, a freely suspended film becomes unstable with respect to arising holes induced by thermal fluctuations. Therefore the films exist in a restricted temperature interval. Another possibility to create very thin films is by using special dopes to conventional soap films the stability of which is supplied by surfactant monolayers on the water surface. Such films may be prepared with the thickness of several  hundred Angstroms and investigated by optical and X-ray methods (see e.g., references in \cite{CH92,KL93}, and more recent papers \cite{OP01,KN01}).

There are different types of orientational ordering in the films. The films can be liquid (isotropic), such films are pulled from a bulk smectic-$A$ phase. Films pulled from low-symmetry smectic phases possess an in-plane orientational order. The films pulled from bulk smectics-$C$ have a ``nematic'' orientational order and ones pulled from bulk hexatic smectics possess a hexagonal orientational order. Here we examine the simplest case of the isotropic films.

We treat the films as two-dimensional systems, i.e., we assume that the thickness of the film is much smaller than the film lateral size. This confines our study to the scales, exceeding the film thickness. Let us stress that in studying freely suspended films there is no interaction with a substrate that is quite important for conventional two-dimensional systems formed on the surface of liquids or solids. Another characteristic peculiarity of such films is a possibility of their bending distortions. If the thickness of the film is sufficiently small fluctuation effects are noticeable on scales exceeding its thickness. The main goal of our work is to investigate the effects.

We start our consideration with a derivation and analysis of the film energy. If the surface tension $\sigma$ of the film is small one should take into account the contribution to the energy related to the film curvature in addition to the contribution from the film surface tension $\sigma$. In the main approximation, the curvature contribution to the film energy can be written as
 \begin{equation}
 {\cal H_\mathrm{s}} = \int dS \left[ \frac{\kappa}{2} (R^{-1}_1
 + R^{-1}_2)^2 + \bar{\kappa}R_1^{-1} R_2^{-1} \right] ,
 \label{memb1}
 \end{equation}
by analogy with the bending energy of lipid bilayers (see the original paper \cite{HE73} and also its textbook version in \cite{SA03}). Here $R_1$ and $R_2$ are local radii of the film and the coefficients $\kappa$ and $\bar\kappa$ are called bending modules (or Helfrich modules). The quantity $R^{-1}_{1} R^{-1}_{2}$ is the Gaussian curvature of the film, and the combination $R^{-1}_{1} +R^{-1}_{2}$ is called its mean curvature. In accordance with the Gauss-Bonnet theorem, the last term in Eq. (\ref{memb1}) (with the coefficient $\bar\kappa$) is the topological invariant. Hence it does not play a role for fluctuations of the film shape that does not change its topology. That is why further (at examining fluctuations effects in the framework of the perturbation series) we ignore the topological contribution to the film energy.

Comparing the curvature energy (\ref{memb1}) with the energy related to the surface tension, one finds the characteristic length $l_\sigma =\sqrt{\kappa/\sigma}$. At scales larger than $l_\sigma$ the surface tension dominates in the film energy. Dynamic properties of the film in the region of scales including fluctuation effects were examined in the work \cite{KL85}. Note to the point that in an experimental setup, where the film area or the film thickness may relax to the equilibrium, the surface tension tends to zero since $\sigma=0$ is just the equilibrium condition. Such experimental setup is realized if the film is suspended without stretching or if the film is connected to a reservoir of molecules constituting the film. Then one expects that there exists the region of scales between the film thickness $l_\mathrm{th}$ and $l_\sigma$ where the bending energy (\ref{memb1}) dominates over the surface tension term. We examine further just the intermediate region of scales. The main goal of our investigation is to examine a role of dynamic fluctuation effects that play a relevant role in the long-scale dynamic properties of the films.

For scales larger than $l_\sigma$ bending distortions of the film propagate, the mode can be termed as bending sound. Its velocity is $c_b=(\sigma/\rho)^{1/2}$ where $\rho$ is the two-dimensional mass density of the film. The bending sound has the dispersion law $\omega= c_b q$, where $\omega$ is frequency and $q$ is wave vector. For scales smaller than $l_\sigma$ the bending distortions still propagate. However, the acoustic dispersion law in this region is substituted by the quadratic dispersion law $\omega =(\kappa/\rho)^{1/2} q^2$. The $q$-dependence of the frequency in this case is similar to one for spin waves in ferromagnets.

As in the case of the bending sound the bending mode with the dispersion law $\omega=(\kappa/\rho)^{1/2} q^2$ has an anomalously weak linear damping, because due to rotational symmetry viscous-like damping is forbidden. The reason is that if the bending viscosity is non-zero it would lead to an energy dissipation in a homogeneously rotating film \cite{KL85,KL93}, that is impossible. That is why the bare bending viscous coefficient is zero and the damping of the bending mode is determined mainly by non-linear fluctuation effects. In our case the main contribution into the damping comes from the interaction of the bending mode with the in-plane viscous mode (unlike crystalline membranes \cite{LK12}, or liquid crystalline films embedded in a viscous liquid \cite{BL75,KL93}).

In the region of negligible surface tension, for scales between $l_\mathrm{th}$ and $l_\sigma$, thermal bending fluctuations of the film shape lead to a logarithmic renormalization of the modules $\kappa$ and $\bar\kappa$. First an attempt to calculate the renormalization of the module $\kappa$ was taken by Helfrich \cite{HE85}, and later by F\"orster \cite{FO86}. The correct renormalization group (RG) equation for the bending module $\kappa$ in the one-loop approximation was derived in papers by Peliti and Leibler \cite{PL85}, Kleinert \cite{KL86a} and Polyakov \cite{PO86}. The RG-equation for $\bar\kappa$ in the same approximation was found by Kleinert \cite{KL86b}. The RG-equation for the bending module $\kappa$ is
 \begin{equation}
 d\kappa/d\xi =-3T/(4\pi),
 \label{filmrg}
 \end{equation}
where $\xi$ is logarithm of the scale, where the modulus is determined, and $T$ is temperature (measured in energy units). The approach implies that $T/\kappa\ll1$.

We see that the dimensionless parameter $T/\kappa$ characterizes intensity of the thermal bending fluctuations. For lipid bilayers, the quantity is usually on the order of $10^{-2}$. The smallness is related to the fact that the bilayer thickness exceeds the atomic length. Therefore one expects that for the freely suspended films the ratio $T/\kappa$ is even smaller since the bending module $\kappa$ is roughly proportional to the third power of the film thickness (as it follows from the classical theory of elastic shells). Further, we treat the ratio $T/\kappa$ as a small parameter. There is another dimensionless parameter characterizing the film, $\kappa\rho/\eta^2$, where $\eta$ is the in-plane shear viscosity coefficient of the film. Like it is in the case for conventional bulk nematics, we expect that the parameter $\kappa\rho/\eta^2$ is small as well.

Let us recall main results of the work \cite{KL85} (see also the monograph \cite{KL93}) concerning the fluctuation damping of the hydrodynamic modes in freely suspended films for the region of scales larger than $l_\sigma$. The bending sound damping comes mainly from thermal fluctuations, the damping can be estimated as $T q^3 /(\rho c_b)$. There is also a fluctuation contribution to the damping of the in-plane sound that can be estimated as
 \begin{equation}
 \left(\frac{T c_b q^5}{\rho}\right)^{1/3}.
 \label{dampold}
 \end{equation}
However, because of a small numerical factor (on the order of $10^{-2}$) in front of the quantity (\ref{dampold}) in the sound damping one should take into account the conventional damping $\sim \eta q^2/\rho$ in addition to the fluctuation contribution. Analogously, one should take into account the conventional damping $\eta q^2/\rho$ of the transverse (to the wave vector $\bm q$) velocity perturbations in addition to the fluctuation contribution (\ref{dampold}) (with a small numerical factor on the order of $10^{-2}$).

Note that the specific entropy $s/\rho$ (where $s$ is the two-dimensional entropy density, $s\, dx \, dy$ is the entropy of the film element) is hardly excited by the fluctuations we are investigating. That is why below the quantity is assumed to be equal to its equilibrium value. Therefore further on all thermodynamic derivatives are assumed to be taken at constant specific entropy $s/\rho$.

We assume that in equilibrium the film is parallel to the $X-Y$ plane and its bending distortions are characterized by the displacement $h$ of the film along the $Z$-direction. The displacement $h$ is treated as a function of $x,y$. Therefore the film shape is determined in the Monge representation as $z=h(x,y)$. Then the area element of the film is expressed as $dS=g^{1/2}\, dx \, dy$, where
 \begin{equation}
 g=1+ (\partial_x h)^2+(\partial_y h)^2.
 \label{stat1}
 \end{equation}
The quantity $g$ plays a role of the determinant of the metric tensor of the film. The unit vector perpendicular to the film has the following components
 \begin{equation}
 l_{i}=g^{-1/2}
 (-\partial_x h,-\partial_y h, 1).
 \label{stat2}
 \end{equation}

The starting point of our analysis is the energy of the film. Further, we treat all variables characterizing the film as functions of $x,y$. Then the first term in the energy (\ref{memb1}) is rewritten as
 \begin{equation}
 {\cal H}_\mathrm{s}
 = \int dx\, dy\, g^{1/2} \frac{\kappa}{2}
 \left[\partial_\alpha (\partial_\alpha h / g^{1/2})\right]^2.
 \label{film1}
 \end{equation}
Here and below Greek subscripts designate components of the vectors along the $X,Y$-axes. In addition to the bending energy (\ref{film1}), we consider also the film energy related to the film compressibility
 \begin{equation}
 {\cal H}_\mathrm{n}=
 \int dx \, dy \,  g^{1/2}\frac{B}{2}
 \left(\frac{\rho}{\rho_0 g^{1/2}}-1\right)^2,
 \label{film2}
 \end{equation}
where $B$ is the compressibility module. Here $\rho_0$ is the ``intrinsic'' equilibrium mass density of the film and $\rho_0 g^{1/2}$ is the equilibrium mass density in projection to the $X-Y$ plane. Note that the expressions (\ref{film1},\ref{film2}) are formally valid for arbitrary variations of the film shape, the only restriction is absence of over-folds, that is uniqueness of the function $h(x,y)$. The contributions (\ref{film1},\ref{film2}) have to be supplemented by the kinetic energy of the film that is written as
 \begin{equation}
 {\cal H}_\mathrm{kin}
 =\int dx\, dy\, \frac{j_x^2+j_y^2+j_z^2}{2\rho},
 \label{filmkin}
 \end{equation}
where $\bm j$ is the film momentum density. Its definition implies that $\bm j\, dx\, dy$ is the momentum of the film element. Then the film velocity is $\bm v=\bm j/\rho$, as usual.

The film surface tension $\sigma$ is a variation of the film energy over the film area. Taking into account that the area element is $g^{1/2}\, dx \, dy$ and varying the expression (\ref{film2}) over $g^{1/2}$, one obtains
 \begin{equation}
 \sigma= -B
 \left[\frac{\rho}{\rho_0 g^{1/2}}-1\right],
 \label{film3}
 \end{equation}
that is an expression of the first order in the film density deviations from its equilibrium value. If the film is stretched then its mass density $\rho$ deviates from the equilibrium value $\rho_0 g^{1/2}$ and the surface tension is non-zero. A non-zero surface tension is produced by fluctuations of $\rho$. Note that just the module $B$ determines the in-plane sound velocity $c_l$: $c_l^2= -\partial\sigma/\partial\rho=B/\rho$.

The dynamic non-linear equations for the film displacement $h$ and for the mass density $\rho$ are
 \begin{eqnarray}
 \rho\partial_t h
 =j_{z}-j_\alpha\partial_\alpha h,
 \label{filmm3} \\
 \partial_t \rho
 =-\partial_\alpha j_\alpha.
 \label{filmm4}
 \end{eqnarray}
The equation (\ref{filmm3}) is the kinematic condition implying that the film moves with the velocity $\bm v =\bm j/\rho$, and the equation (\ref{filmm4}) is the mass conservation law. Therefore the equations (\ref{filmm3},\ref{filmm4}) are exact that is there are no dissipative corrections to the equations.

The equations (\ref{filmm3},\ref{filmm4}) should be supplemented by the equation for the momentum density $j_i$ of the film that is written as the momentum conservation law
 \begin{equation}
 \partial_t j_i = -\partial_\alpha\left(v_\alpha j_i
 + g^{1/2}{\cal T}_{i\alpha}
 -g^{1/2}\eta_{i\alpha \beta m}\partial_\beta v_{m}\right),
 \label{film15}
 \end{equation}
where Latin subscripts run over $x,y,z$. All quantities in the equations are, as above, assumed to be functions of $x,y$. The expression for the stress tensor ${\cal T}_{ik}$ was found in the work \cite{LM89} (see also the book \cite{KL93}), it is written as
 \begin{eqnarray}
 {\cal T}_{ik} =  - \left[\sigma
 + \frac{\kappa}{ 2}(\nabla \bm l)^{2}\right] \delta^{\perp}_{ik}
 + \kappa(\nabla \bm l) \partial^{\perp}_k l_i
 - \kappa l_i \partial^{\perp}_k \nabla \bm l.
 \label{filmm6}
 \end{eqnarray}
Here $\delta^\perp_{ik}\equiv\delta_{ik}-l_i l_k$ stands for the projector to the film, the unit vector $\bm l$ has the components (\ref{stat2}), $\nabla\bm l=\partial_\alpha l_\alpha$, and $\partial^\perp_k=\delta^\perp_{k\alpha}\partial_\alpha$. The viscosity  tensor can be written as
 \begin{eqnarray}
 \eta_{iklm}=(\eta-\zeta)
 \delta^{\perp}_{ik}\delta^{\perp }_{lm}
 +\eta(\delta ^{\perp }_{il}\delta^{\perp}_{km}
 +\delta ^{\perp }_{im}\delta ^{\perp }_{kl}) .
 \label{filmm19}
 \end{eqnarray}
Here $\eta$ and $\zeta$ are two-dimensional analogs of the three-dimensional first (shear) and second (bulk) viscosity coefficients.

In the linear approximation (for the region of scales under consideration, smaller than $l_\sigma$) one finds the following hydrodynamic (i.e., long-scale) modes of the film:
\begin{itemize}
\item
the in-plane propagating sound mode with the linear dispersion law $\omega = c_l q$;
\item
the overdamped in-plane thermodiffusion mode;
\item
the overdamped in-plane viscous mode;
\item
the bending (flexular) propagating mode with the dispersion law $\omega = \sqrt{\kappa/\rho}\ q^2$.
\end{itemize}
The specific entropy dynamics (thermodiffusion mode) can be separated from the other modes. Then, in what follows we assume, that the specific entropy mode is not excited, and therefore one can safely forget about the thermodiffusion. Note that in the linear approximation the in-plane sound mode has the standard viscous damping $\sim (\eta/\rho)q^2$, whereas for the bending mode such kind of viscous damping is forbidden by the rotational symmetry of the film (for details see \cite{KL93}). Only higher order over the gradients (e.g., proportional to $q^4$) dissipative terms are not forbidden, and such terms will produce a non-zero damping of the bending mode. However, as we demonstrate below, thermal fluctuations produce a larger contribution to the bending mode damping, that is proportional to $q^2$. Calculation of the damping is one of the main goals of our work.

In the hydrodynamic regime (that is for scales, larger than the film thickness) the in-plane sound is harder, i.e., its frequency $\omega$ for a given wave vector ${\bf q}$ is much higher than the characteristic frequencies (inverse time scales) for the bending and for the in-plane viscous modes. Therefore the degrees of freedom related to the sound (density fluctuations and the longitudinal to the wave vector component of $j_\alpha$) can be effectively excluded from the consideration. Then for the first step of the analysis we stay with the closed description of the bending and of the shear viscous modes. We analyze their non-linear interaction leading, particularly, to fluctuation contributions to damping of the modes.

Particularly, we establish, that the non-linear fluctuation effects produce the viscous-like damping of the bending mode that leads to the following dispersion law
 \begin{equation}
 \omega=\sqrt{\kappa/\rho}\ q^2-i \nu q^2/2,
 \label{filmdisp}
 \end{equation}
where $\nu$ is the fluctuation bending viscosity. The second term in the right-hand side of Eq. (\ref{filmdisp}) gives the bending mode damping related to scattering of the bending mode on fluctuations of the viscous in-plane mode. That is why $\nu$ is proportional to the temperature $T$:
 \begin{equation}
 \nu=\frac{T\eta}{\pi\kappa\rho}.
 \label{film35}
 \end{equation}
A derivation of the expression (\ref{film35}) is given in the next sections.

 \section{Effective action}
 \label{sec:action}

We proceed to calculating the bending mode damping (\ref{filmdisp},\ref{film35}) and the corrections to the shear viscosity $\eta$ caused by thermal fluctuations. The calculations can be done in the framework of the diagrammatic technique first developed by Wyld \cite{WY61} in the framework of hydrodynamic turbulence. Then the technique was generalized for a wide class of systems by Martin, Siggia and Rose \cite{MS73}. Note that the technique is the classical limit of the Keldysh diagrammatic technique \cite{Ke64} (see also the textbook \cite{Kinetics}). We use the version proposed by Janssen \cite{JA76}, where correlation functions of the fluctuating fields are written as path (functional) integrals over the fields with the weight $\exp(iI)$ where $I$ is an effective action constructed in accordance with the dynamical equations of the system (see also the book \cite{KL93}).

The effective action $I$ for the problem we are investigating is the sum $I=I_\mathrm{reac}+I_\mathrm{diss}$ of the reactive and of the dissipative contributions,
 \begin{eqnarray}
 I_\mathrm{reac}=\int dt\, dx \, dy \, \bigl[
 \mu ( \rho \partial_t h -j_z +j_\alpha \partial_\alpha h)
 \nonumber \\
 +p_i \partial_t j_i+ p_i\partial_\alpha(v_\alpha j_i)
 +p_i \partial_\alpha ( \sqrt g\, {\cal T}_{i\alpha}) \bigr],
 \label{raction} \\
 I_\mathrm{diss}=\int dt\, dx \, dy \,  \sqrt g\,
 \partial_\alpha p_i
 \eta_{i\alpha \beta m}\left[\partial_\beta v_{m}
 +iT \partial_\beta p_{m} \right],
 \label{daction}
 \end{eqnarray}
where $\mu$ and $\bm p$ are auxiliary fields. Here we omitted the specific entropy fluctuations. The action $I$ is constructed in accordance with the dynamical equations (\ref{filmm3}) and (\ref{film15}), and the relation (\ref{filmm4}) is implied. The last term in Eq. (\ref{daction}) comes from viscous Langevin (random) forces. That is why the term is proportional to the temperature $T$. Note that the action (\ref{raction}-\ref{daction}) is formally exact in $h$.

Note that the pair correlation functions, like $\langle h \mu \rangle$ (here and henceforth angular brackets mean averaging that is equivalent to a functional integral with the weight $\exp(iI)$) determine response of the system to an external excitation (force). The structure of the action (\ref{daction}) guarantees validity of the fluctuation-dissipation theorem. The theorem leads to the relation between the correlation functions like $\langle h h\rangle$ and $\langle h \mu\rangle$. Note also, that the pair correlation functions like $\langle \mu \mu\rangle$ are zero.

Both, the bending and the in-plane viscous modes have the bare dispersion laws $\omega\propto q^2$. Therefore in the long-scale limit (small wave vectors) in the main approximation one can neglect the time derivative in the mass conservation law (\ref{filmm4}) to obtain the condition $\partial_\alpha j_\alpha=0$. It is analogous to the incompressibility condition for the Navier-Stokes equation.
Next, an integration over the density field $\rho$ (that is the variable related to the in-plane sound) at calculating the correlation functions can be performed in the saddle-point approximation. By other words, the action $I$ can be substituted by its saddle-point value that can be found by equating to zero the variation of the action (\ref{raction}) over $\rho$. The main $\rho$-dependence of the action is related to the contribution to the stress tensor ${\cal T}_{i\alpha}$, proportional to the surface tension $\sigma$, see the expressions (\ref{film3},\ref{filmm6}). Calculating the variation of the action (\ref{raction}) over $\rho$ and equating the variation to zero, one obtains the following condition
 \begin{equation}
 \delta^\perp_{\alpha i} \partial_\alpha p_i=0.
 \label{filmpc}
 \end{equation}
Particularly, the condition (\ref{filmpc}) leads to the conclusion that the term with the surface tension and the term with the difference $\eta-\zeta$ in the viscosity tensor (\ref{filmm19}) do not enter the effective action for the interacting bending and viscous modes.

In this approximation, the surface tension $\sigma$ is not a dynamic variable, like pressure in the Navier-Stokes equation. The quantity $\sigma$ is passively followed to the bending and the viscous fluctuations. Thus, variations of the surface tension are relatively weak. Therefore, to determine the mass density, the surface tension $\sigma$ can be put to zero. Then, we obtain from Eq. (\ref{film3}) that
 \begin{equation}
 \rho=\rho_0 \sqrt{1+(\nabla h)^2},
 \label{filmrho}
 \end{equation}
where $\rho_0$ is the ``internal'' mass density of the film. The expression (\ref{filmrho}) leads to the following renormalization law
 \begin{equation}
 \frac{d\ln\rho_0}{d\xi}=
 \frac{T}{4\pi\kappa},
 \label{filmrhor}
 \end{equation}
where $\xi$ is logarithm of the current scale.

One should choose variables describing the bending and the viscous modes. We take the physical variables $h, j_z, j_\alpha$. The last one satisfies the condition $\partial_\alpha j_\alpha=0$, therefore we deal with three scalar fields. As to the auxiliary fields, we choose $\mu$, $p_z$ and the transverse (to the wave vector) component $p^{tr}_\alpha$ of $p_\alpha$, satisfying the condition $\partial_\alpha p^{tr}_\alpha=0$.

In the framework of the perturbation theory one starts from the ``bare'' correlation functions that can be easily obtained by Gaussian integration if we keep solely the second-order term in the effective action $I$. Explicit expressions for the correlation functions of the viscous mode are given by the expressions
 \begin{eqnarray}
 \langle j_\alpha(t,\bm r) j_\beta(0,0) \rangle
  =\int \frac{d\omega\ d^2q}{(2\pi)^2}
 e^{-i\omega t+i \bm q \bm r}F_{\alpha\beta},
 \nonumber \\
 F_{\alpha\beta}=
 \frac{2T\eta(q^2\delta_{\alpha\beta}-q_\alpha q_\beta)}{\omega^2+\eta^2 q^4 /\rho^2},
 \label{film23} \\
 \langle j_\alpha(t,\bm r) p^{tr}_\beta(0,0)\rangle
 =\int \frac{d\omega\ d^2q}{(2\pi)^2}
 e^{-i\omega t+i \bm q \bm r} G_{\alpha\beta},
 \nonumber \\
 G_{\alpha\beta}=-\frac{\delta_{\alpha\beta}-q_\alpha q_\beta/q^2}
 {\omega+i\eta q^2/\rho}.
 \label{film24}
 \end{eqnarray}
However, to analyze  fluctuation effects one should use ``dressed'' correlation functions of $h$, $j_z$, i.e., with included fluctuation contributions. Then the pair correlation function of the displacement $h$ is written as
 \begin{eqnarray}
 \langle h(t, \bm r) h(0,0) \rangle
 =\int \frac{d\omega\ d^2q}{(2\pi)^2}
 \exp(-i\omega t+i \bm q \bm r)F_{hh},
 \nonumber \\
 F_{hh}=\frac{2 \Pi(\omega,\bm q)}
 {[\omega^2 -\kappa q^4/\rho+ \Sigma(\omega)]
 [\omega^2 -\kappa q^4/\rho+\Sigma(-\omega)]},
 \label{filmbe}
 \end{eqnarray}
where $\Pi$ and $\Sigma$ are ``polarization'' and ``self-energy'' functions, in the terminology borrowed from the quantum field theory. The quantities $\Pi$ and $\Sigma$ have to be calculated in the framework of the perturbation theory. Due to the fluctuation-dissipation theorem
 \begin{equation}
 \mathrm{Im}\, \Sigma(\omega,q)
 =-\mathrm{Im}\, \Sigma(-\omega,q)
 =\frac{\rho\omega}{T}\Pi(\omega,q).
 \label{filmFDT}
 \end{equation}

One finds that the main contributions to $\Pi$ and $\Sigma$ are proportional to $q^2$. The real part of $\Sigma$ reproduces the renormalization of $\kappa$ and $\rho_0$ that is irrelevant for us due to the inequality $T/\kappa\ll1$. That is why further we take into account only the imaginary part of $\Sigma$: $\mathrm{Im}\, \Sigma=\nu q^2$, $\Pi=(T/\rho) \nu q^2$. In the approximation we obtain from Eq. (\ref{filmbe})
 \begin{eqnarray}
 \langle h h \rangle
 =\int \frac{d\omega\ d^2q}{(2\pi)^2}
 \exp(-i\omega t+i \bm q \bm r)F_{hh},
 \nonumber \\
 F_{hh}=\frac{2T\rho^{-1}\nu q^2}
 {(\omega^2 -\kappa q^4/\rho)^2+\nu^2 q^4 \omega^2}.
 \label{film22}
 \end{eqnarray}
The expressions correspond to the dispersion law (\ref{filmdisp}).

\section{Perturbation theory}
\label{sec:perturbation}

In this section we calculate the non-linear fluctuation effects in the framework of the perturbation theory. For the purpose we expand the effective action $I$. We use the one-loop approximation that is one should expand the effective action up to the fourth order. Results of the calculations in the framework of the perturbation theory can be presented by Feynman diagrams, where lines correspond to the pair correlation functions (\ref{film23},\ref{film24},\ref{film22}) and vertices are determined by the third and fourth order terms in the effective action. Note that the expansion of the contributions to the action (\ref{raction},\ref{daction}) has to be performed in $h$ since the terms are quadratic in the other fields. One should expand in $h$ the relation (\ref{filmpc}) as well. The expansion determines the longitudinal (to the wave vector) component of $p_\alpha$. We find
 \begin{equation}
 \partial_\alpha p_\alpha=
 -\partial_\alpha h \partial_\alpha p_z+
 \partial_\alpha h  \partial_\beta h \partial_\alpha p^{tr}_\beta,
 \label{filmpc1}
 \end{equation}
with the second-order accuracy in $h$.

Now we proceed to calculate the fluctuation contribution to the polarization function $\Pi$ entering the expression (\ref{filmbe}). Because we assume $\eta ^2 \gg \kappa \rho$ the leading interaction vertices are determined by solely the dissipative effective action (\ref{daction}). To find $\Pi$ one should select the terms with the field $p_z$. All relevant third order terms containing $p_z$ are
 \begin{eqnarray}
 I_\mathrm{third}
 =\int dt\, dx \, dy \, \biggl\{
 \frac{\eta}{\rho}\partial_\beta h
 \partial_\alpha p_z
 (\partial_\beta j_\alpha
 +\partial_\alpha j_\beta)
 \nonumber \\
 +2iT\eta\partial_\beta h
 \partial_\alpha p_z
 (\partial_\beta p^{tr}_\alpha
 +\partial_\alpha p^{tr}_\beta)\biggr\}.
 \label{film31}
 \end{eqnarray}
The fourth order term needed for us is
 \begin{eqnarray}
 I_\mathrm{fourth}= \int dt\, dx \, dy \,
 \biggl\{2i T\eta \partial_\mu h \partial_\nu h
 \partial_\mu p_z \partial_\nu p_z
 \nonumber \\
 -4iT\eta\partial_\mu p_z \partial_\mu h
 \frac{\partial_\alpha \partial_\beta}{\nabla^2}
 (\partial_\beta h \partial_\alpha p_z)
 \nonumber \\
+iT\eta[\partial_\beta h\partial_\alpha h
 +\delta_{\alpha \beta} (\nabla h)^2]
 \partial_\alpha p_z \partial_\beta p_z \biggr\},
 \label{film30}
 \end{eqnarray}
it is quadratic in $p_z$. Note that the action (\ref{film30}) is non-local (it contains the term proportional to $1/\nabla^2$), as one could expect because this action is derived by the integration-out of the full effective action over the acoustic degrees of freedom.

 \begin{equation}
 \Pi_1=\feyn{x f0 gl g3 g4}
 \label{feyn1}
 \end{equation} \vspace{.5cm}

 \begin{equation}
 \Pi_2 = \feyn{f gl f}
 \label{feyn2}
 \end{equation}

 \begin{equation}
 \Pi_3=\feyn{f gl h}
 \label{feyn3}
 \end{equation}

There are some one-loop contributions to the quantity $\Pi$, that can be represented by the Feynman diagrams, see Eqs. (\ref{feyn1}-\ref{feyn3}). The cross in Eq. (\ref{feyn1}) represents the fourth-order vertex determined by the action (\ref{film30}) and the wavy line stands for the pair correlation function (\ref{film22}). The third-order vertices in Eqs. (\ref{feyn2}-\ref{feyn3}) are determined by the action (\ref{film31}). The solid line there represents the pair correlation function (\ref{film23}) and the combined solid-dashed line represents the pair correlation function (\ref{film24}). The diagrams enable one to write an explicit expression for $\Pi$.

The structure of the vertices leads to the conclusion that $\Pi(k)\propto k^2$, where $k$ is wave vector. Substituting $\Pi=(T/\rho) \nu k^2$ one finds after lengthy but straightforward mathematics
 \begin{eqnarray}
 \frac{\nu\kappa\rho}{T\eta}=5\frac{k_\alpha k_\beta}{k^2}\int \frac{d^2q}{(2\pi)^2}
 \frac{(q_\alpha+k_\alpha)(q_\beta+k_\beta)}{ |\bm q+\bm k|^4}
 \nonumber \\
 -4\frac{k_\alpha k_\mu}{k^2}
 \int \frac{d^2q}{(2\pi)^2}\frac{(q_\mu +k_\mu)(q_\beta+k_\beta)}{ |\bm q+\bm k|^4}
 \frac{q_\alpha q_\beta}{ q^2}
 \nonumber \\
 - \frac{k_\beta k_\nu}{k^2}
 \int \frac{d^2q}{(2\pi)^2}
 \frac{ (q_\alpha+k_\alpha)(q_\mu+k_\mu)}
 { |\bm q+\bm k|^4}\frac{1}{q^4}
 \nonumber \\
 \{q_\alpha q_\nu (q^2\delta_{\beta\mu} -q_\beta q_\mu)
 +q_\beta q_\nu (q^2\delta_{\alpha\mu} -q_\alpha q_\mu)
 \nonumber \\
 +q_\alpha q_\mu (q^2\delta_{\beta\nu} -q_\beta q_\nu)
 +q_\beta q_\mu (q^2\delta_{\alpha\nu}-q_\alpha q_\nu )\}.
 \label{film32}
 \end{eqnarray}
Simple inspection of this expression shows that all the integrals converge in the both limits, at large $q$ and at small $\bm q+\bm k$. Therefore the main contribution into the integral comes from the region $q\sim k$ and there are no ultraviolet contributions (particularly, logarithms) to the quantity $\nu$, in accordance with the symmetry arguments. The dimension analysis leads to the conclusion that the quantity (\ref{film32}) is independent of $\bm k$. Calculations (collected in Appendix) give the final answer (\ref{film35}).

Out of these calculations will come a bit unexpected result, that forbidden by rotational symmetry viscous-like damping of the bending mode, becomes perfectly legitimate driven by thermal fluctuations. Note to the point, that the situation where the fluctuation damping (determined by fluctuations of the same scale as the wavelength) has some scaling forbidden for the bare damping by symmetry is characteristic also of the spin waves in two-dimensional ferromagnets \cite{PF77}.

Let us estimate a fluctuation correction to the shear viscosity coefficient $\eta$. As previously, we consider only the interaction terms produced by the dissipative part of the effective action (\ref{daction}), that is justified by the inequality $\kappa\rho\ll \eta^2$. One obtains the principal interaction term
 \begin{eqnarray}
 I_\mathrm{int}=\int dt\, dx \, dy \, [
 (\eta/\rho)\partial_\beta h
 \partial_\alpha j_z
 (\partial_\beta p^{tr}_\alpha
 +\partial_\alpha p^{tr}_\beta)
 \nonumber \\
 +2iT\eta\partial_\beta h
 \partial_\alpha p_z
 (\partial_\beta p^{tr}_\alpha
 +\partial_\alpha p^{tr}_\beta)].
 \label{filmint}
 \end{eqnarray}
The one-loop contribution to $\eta$ is determined by the same diagrams (\ref{feyn2}-\ref{feyn3}) where interaction triple vertices are determined now by the action (\ref{filmint}). Calculations lead to the following estimation of the fluctuation correction
 \begin{equation}
 \Delta \eta \sim \frac{T}{\kappa}
 \frac{\eta^2}{\sqrt{\kappa\rho}}.
 \label{filmcorr}
 \end{equation}

Now we can accurately formulate the applicability conditions of our approach. Our procedure is correct provided $\Delta\eta\ll \eta$ that is if
 \begin{equation}
 \epsilon\equiv\frac{T}{\kappa}
 \frac{\eta}{\sqrt{\kappa\rho}}\ll 1.
 \label{filmappl}
 \end{equation}
Note that the condition (\ref{filmappl}) implies that the damping of the bending mode is much less than its frequency since the inequality (\ref{filmappl}) is equivalent to the condition $\nu\ll \sqrt{\kappa/\rho}$, see Eq. (\ref{filmdisp}).

As a note of caution we should say also, that above we neglected the self-interaction of the viscous mode that leads to the logarithmic corrections to the shear viscosity coefficient $\eta$ that can be estimated as \cite{Forster-Nelson-Stephen,Andreev}
 \begin{equation*}
 \Delta\eta \sim \rho T/\eta,
 \end{equation*}
up to a logarithmic factor. Comparing the estimation with Eq. (\ref{filmcorr}) we conclude that the effect we calculated above is stronger due to the inequality $\eta^2 \gg \kappa\rho$.

It is interesting that just the parameter (\ref{filmappl}) is the small parameter justifying the perturbation expansion in $h$. If the parameter is not small then we are beyond the applicability of the perturbation theory, in the region of strong interaction. Then nothing can be done in the framework of the perturbation theory. Based on heuristic hand-waving arguments related to the fluctuation enhancement of the viscosity coefficient we could only speculate that probably, the behavior in that region corresponds to a glass state of the film since an essential viscosity growth is expected in the case.

\section{Conclusion}
\label{sec:conclusion}

In conclusion, in this paper we describe dynamic fluctuation phenomena in freely suspended films. We restrain ourselves to liquid-like (isotropic) films in the overlooked in the previous works conditions (tensionless membrane freely suspended in the gas or vacuum). The bending (flexular) mode of such film turns out to be soft and weakly attenuated, therefore yields to strong dynamic fluctuation effects. In the harmonic approximation, for the bending mode there is no viscous-like, proportional to $q^2$ ($q$ is the wave vector of the mode) attenuation, and only much smaller super-viscous attenuation $\propto q^4$ is not forbidden by the rotational symmetry. We calculate the dominant fluctuation contributions to the damping of the bending mode due to its coupling to the viscous (non-propagating) in-plane mode. The fluctuation damping restores the viscous-like $q^2$ attenuation law of the bending mode. The damping is weak due to smallness of the dimensionless coupling constant obligatory within our perturbation approach. What is fascinating about our results is that they not only contribute to understanding of many dynamic properties important for nano-technological opto-mechanical applications of the films, the results can be confronted with recent experimental data \cite{HW13} on fluctuation enhancement of the membrane viscosity. Another way to check experimentally our theory predictions is to perform inelastic (dynamic) light scattering experiments (like in the works \cite{OP01,KN01}) to measure the scattered signal (intensity) line-width.

What makes our calculations involved and non-trivial, is the fact that there are two essential dimensionless parameters in the theory, namely, $T/\kappa \ll 1$ and $\eta ^2/(\kappa \rho ) \gg 1$, and physics depends crucially on how they interplay. The small parameter of our perturbation theory is $\epsilon$, see Eq. (\ref{filmappl}). One can try to extract the value of $\epsilon$ from literature data on material properties of the lipid membranes and the liquid crystals that are rather dispersed, see, e.g., \cite{materials}. There are some materials where (at room temperatures!) $\epsilon < 1$, and also those with $\epsilon >1$, but the both inequalities are not too strong. One also should keep in mind that there is regular temperature dependence (a sort of Arrhenius law) of the bare material parameters \cite{materials}. If we stretch all essential material parameters to their utmost (but still not unrealistic) values, the ratio could be either $\epsilon \ll 1$, or $\epsilon \gg 1$. If the ratio is not small then we go beyond the applicability of the perturbation theory, into the region of strong interaction. This problem deserves a separate investigations. In this regime of very strong bending fluctuations one can expect the membrane to be broken via non-perturbative processes, like pore formation. Another theoretically tempting possibility would be vitrification of the membrane.

We believe that our results are also a step forward in establishing connections between different regimes of the collective membrane
dynamics. In own turn thinking about membrane dynamics from a pure physics-based ground can bring new insight on some relevant characteristics of biophysical cells. We expect fluctuation dynamics of freely suspended membranes to be a fascinating and fruitful field of investigations for biologists, applied scientists, and experimental and theoretical physicists.

\acknowledgments

Our work was funded by Russian Science Foundation (grant 14-12-00475).

\appendix

 \section{}
 \label{app1}

To calculate the integral (\ref{film32}) we substitute there $\bm k=(1,0)$ to obtain
 \begin{eqnarray}
 \frac{\nu\kappa\rho}{T\eta}=
 \int \frac{d\varphi\ dq\ q }{(2\pi)^2}
 \frac{1}{(1+q^2 -2q \cos\varphi)^2} \qquad
 \label{film33} \\
 \left[ q^2 \cos(2\varphi) -2q \cos\varphi
 +\frac{5}{2} -2 \cos(2\varphi) +\frac{1}{2}\cos(4\varphi) \right].
 \nonumber
 \end{eqnarray}
The integral over the angle $\varphi$ can be found using the following relations:
 \begin{eqnarray}
 \frac{1}{2\pi} \int_0^{2\pi}
 \frac{d\varphi\ \cos(m\varphi)}{(1+q^2-2q \cos\varphi)^2}
 \nonumber \\
 = \frac{(m+1)q^m+(1-m)q^{2+m}}{(1-q^2)^3},
 \nonumber
 \end{eqnarray}
if $0<q<1$, $m\geq 0$, and
 \begin{eqnarray}
  \frac{1}{2\pi} \int_0^{2\pi}
 \frac{d\varphi\ \cos(m\varphi)}{(1+q^2-2q \cos\varphi)^2}
 \nonumber \\
 = \frac{(m+1)q^{2-m}+(1-m)q^{-m}}{(q^2-1)^3},
 \nonumber
 \end{eqnarray}
if $q>1$, $m\geq 0$. Thus the integral over $q$ is divided into two intervals: from $0$ to $1$ and from $1$ to $\infty$. The integration can be easily performed and one finds
 \begin{eqnarray}
 \frac{\nu\kappa\rho}{T\eta}=
 \int_0^1 \frac{dq\ q}{2\pi}\frac{1}{(1-q^2)^3}\frac{5}{2}(1-q^2)^3
 \nonumber \\
 +\int_1^\infty \frac{dq\ q}{2\pi}\frac{1}{(q^2-1)^3}\frac{3}{2q^4}(q^2-1)^3=\frac{1}{\pi}.
 \label{film34}
 \end{eqnarray}
Thus we arrive at the result (\ref{film35}).

\end{document}